\documentclass[aps,prl,twocolumn,showpacs,superscriptaddress]{revtex4}

\usepackage{amssymb}
\usepackage{graphicx}
\usepackage{color}

\begin{document}

\title{Circular-Polarization-Dependent Study of Microwave-Induced
Conductivity Oscillations in a Two-Dimensional Electron Gas on Liquid Helium}

\author{A.~A. Zadorozhko}
\affiliation{Okinawa Institute of Science and Technology, Tancha 1919-1, Okinawa 904-0495, Japan}
\author{Yu.~P. Monarkha}
\affiliation{Institute for Low Temperature Physics and Engineering, 47 Nauky Avenue,
61103, Kharkiv, Ukraine}
\author{D. Konstantinov}
\email[E-mail: ]{denis@oist.jp}
\affiliation{Okinawa Institute of Science and Technology, Tancha 1919-1, Okinawa 904-0495, Japan}
\date{\today}

\begin{abstract}
The polarization dependence of photoconductivity response at
cyclotron-resonance harmonics in a nondegenerate two-dimensional (2D)
electron system formed on the surface of liquid helium is studied using a
setup in which a circular polarization of opposite directions can be
produced. Contrary to the results of similar investigations reported for
semiconductor 2D electron systems, for electrons on liquid helium, a strong
dependence of the amplitude of magnetoconductivity oscillations on the
direction of circular polarization is observed. This observation is in
accordance with theoretical models based on photon-assisted scattering and,
therefore, it solves a critical issue in the dispute over the origin of
microwave-induced conductivity oscillations.
\end{abstract}


\maketitle

\indent Studies of microwave (MW) photoconductivity in a two-dimensional (2D)
electron gas of semiconductor heterostructures subjected to a perpendicular
magnetic field $B$ have revealed remarkable magnetotransport phenomena:
giant microwave-induced resistance oscillations (MIRO)~\cite{ZudDuSim-2001,YeEngTsu-2001}
and associated zero resistance states (ZRS)~\cite{ManSmeKlit-2002,ZudDuPfe-2003}.
These discoveries have opened a prominent research area and triggered
a large body of theoretical works. The universality of the effect of MIRO was
proved by similar observations in hole systems~\cite{ZudMirEbn-2014} and in a
nondegenerate 2D electron system formed on the free surface of liquid helium~\cite{YamAbdBad-2015}.

It is very surprising that by now there is a great body of different
theoretical mechanisms explaining MIRO which use quantum and classical
effects (for a review, see~\cite{DmiMirPol-2012}), but the origin of these
oscillations is still under debate. Among these mechanisms there is a large
group of models whose description is based on the concept of the
photon-assisted scattering off disorder which overcomes the selection rules
existing for direct photon-induced transitions; direct transitions can be
only between adjacent Landau levels: $n^{\prime }-n=\pm 1$ . The
photon-assisted scattering leads to two important effects. Firstly, it gives
a separate contribution to magnetoconductivity $\sigma _{xx}$ where the
displacement of the electron orbit center $X^{\prime }-X$ caused by energy
conservation changes its sign when the ratio $\omega /\omega _{c}$ passes an
integer $m=n^{\prime }-n=1,2,...$ (here $\omega $ is the MW frequency and $%
\omega _{c}$ is the cyclotron frequency)~\cite{Ryz-1969,DurSachRea-2003}. This
mechanism is called the displacement model (DM).

Secondly, electron
scattering to higher Landau levels ($n^{\prime }-n=2,3...$) changes the distribution
function of electrons at these levels $f\left( \varepsilon \right) $: it
acquires an oscillatory form with maxima, and,
therefore, a sort of population inversion occurs~\cite{DmiMirPol-2003,DmiVavAle-2005}.
This mechanism is called the inelastic model (IM) because the inelastic thermalization
rate is an important quantity for the description of the effect. The maxima of
$f\left( \varepsilon \right) $ affect the contribution to $\sigma _{xx}$ caused by
usual scattering processes and also lead to a sign-changing correction to $%
\sigma _{xx}$.

Both theoretical models (DM and IM) give satisfactory descriptions of
MIRO in semiconductor electron systems. Still, there is a critical
unresolved issue which concerns the dependence of MIRO on the direction of
circular polarization: the results of the DM and IM are very
sensitive to the direction of circular polarization, while the MIRO observed
in semiconductor heterostructures are notably immune to the sense of
circular polarization~\cite{SmetGorJia-2005}, or have a very weak dependence
on the direction of circular polarizations in the THz range~\cite{HerDmiKoz-2016}
which is at odds with existing theories of MIRO.

The theoretical analysis~\cite{Mon-2017} of the DM and IM performed for a nondegenerate 2D
electron gas on liquid helium also indicates that the both models result in
practically the same strong dependence of the amplitude of MIRO on the
direction of circular polarization if the number $m=2,3,...$ is not large.
Since the MIRO observed for electrons on liquid helium~\cite{YamAbdBad-2015}
are in good accordance with the IM, and the recent observation of MW-induced
oscillations in magnetocapacitance of a semiconductor system~\cite{DorKapUma} also supports the
concept of a nonequilibrium distribution function oscillating with energy,
there is a strong need to investigate the dependence of the amplitude of
magnetoconductivity oscillations on the direction of circular polarization
for the 2D electron system on liquid helium.

Here, we report the first observation of a strong dependence of the
amplitude of microwave-induced magnetoconductivity oscillations on the direction of
circular polarization in the electrons-on-helium system. The analysis of data given here
shows that this observation is in accordance with theoretical models based on
photon-assisted scattering. Thus, contrary to the mysterious contradiction between experiment
and theory existed in semiconductor heterostructures,
the circular-polarization-dependent study of MIRO in a nondegenerate 2D electron gas
on liquid helium provides strong support for the photon-assisted scattering as the origin
of the magnetoconductivity oscillations.

The experiments are done in a 2D electron system formed on the free surface of liquid $^3$He,
which is contained in a closed cylindrical cell and cooled to $T=0.2$~K. At this temperature the
scattering of electrons is dominated by capillary-wave excitations of the liquid helium surface
(ripplons) and is very well understood~\cite{Andrei_book, Monarkha_book}.
The magnetic field $B$ is applied perpendicular to the liquid surface, and
the longitudinal conductivity of electrons $\sigma_{xx}$ is measured by the capacitive-coupling
method using a pair of gold-plated concentric circular electrodes
(Corbino disk) placed beneath and parallel to the liquid surface. The electrodes have outer diameters
of 14 and 19.8~mm and are separated by a gap of width $0.005$~mm. Similar to our previous experiment~\cite{YamAbdBad-2015},
conductivity oscillations are excited by the electric field component of the fundamental
TEM$_{002}$ mode in a semiconfocal Fabry-Perot resonator~\cite{Kog-1966}. The resonator is formed by
the Corbino disk acting as a flat reflecting mirror and a copper concave (radius of curvature 30~mm)
mirror of diameter 35.2~mm placed above and parallel to the Corbino disk at a distance about 13~mm.
At liquid helium temperatures, the TEM$_{002}$ mode has a frequency $\omega_r/2\pi\approx 35.21$~GHz and
the quality factor is about 10$^4$.

The MW excitation is supplied from a room-temperature source followed by a linear-to-circular polarization
converter, transmitted into the cryostat via a circular (inner diameter 6.25~mm) 1.5 meter long waveguide,
and then coupled to the resonator via a circular aperture (diameter 1.8 mm) drilled in the center of the
concave mirror. The axial symmetry and alignment of this setup are very important to ensure the circular
polarization of the resonant mode field excited in the resonator. To check the latter, we observed the
dependence of the cyclotron resonance (CR) excited in 2D electron system by the circular-polarized MWs on
the direction of the applied magnetic field. For this, the electrons were placed in the maximum of the MW
$E$-field in the resonator by adjusting the height of liquid helium to be about 2.1~mm above the Corbino disk,
and the value of the magnetic field $B$ was adjusted such that the cyclotron frequency of electrons, $\omega_c=eB/m_e c$,
was close to the frequency of the resonant mode.

\begin{figure}[tbp]
\begin{center}
\includegraphics[width=8.5cm]{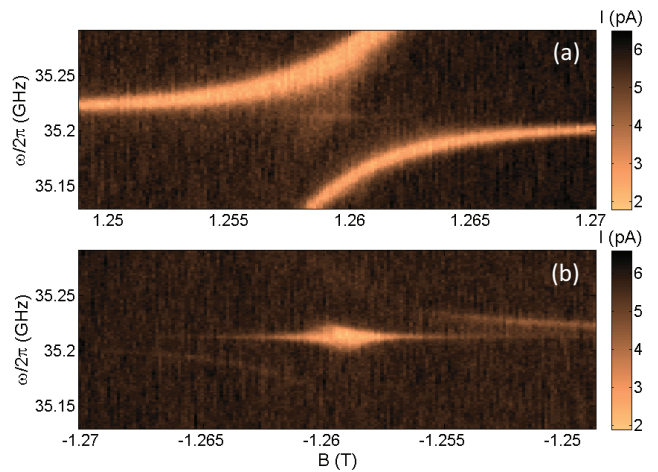}
\end{center}
\caption{(color online) Photocurrent response of the 2D electron system at $T=0.2$~K
and electron density $n_s= 8.2\times 10^7$~cm$^{-2}$ to a circular-polarized microwave excitation
plotted as a function of MW frequency $\omega$ and perpendicular magnetic field $B$.
Panels (a) and (b) are for two opposite directions of the field distinguished by the sign of $B$.}
\label{fig:1}
\end{figure}

To detect the photoconductivity response of electrons we applied
a 20~mV excitation voltage at the frequency of 1.117~kHz to the inner electrode of the Corbino disk and measured
the current induced in the outer electrode by the electron motion. This current is plotted in Fig.~\ref{fig:1}
as a function of the magnetic field $B$ and frequency of MW excitation $\omega$.
Two panels correspond to two opposite directions of the magnetic field $B$. For a given direction of
circular polarization, strong CR absorption should occur only for the proper direction of the magnetic field and
should be strongly suppressed for the opposite direction. As was demonstrated earlier~\cite{AbdYamBad-2016},
for a resonator of sufficiently high quality factor the strong coupling of cyclotron motion of electrons to
the resonator mode leads to the appearance of two polaritonic branches of coupled electron-mode motion. This is
shown in Fig.~\ref{fig:1}(a) where two polaritonic branches are revealed in the conductivity response of
electrons due to their strong heating by the CR absorption. For the opposite direction of the magnetic field, see
Fig.~\ref{fig:1}(b), the polaritonic branches are barely visible, pointing out that the CR absorption is
strongly suppressed. The origin of a strong response appearing around the crossover point $\omega_r=\omega_c$,
which corresponds to $B\approx 1.26$~T, is not clear. For the opposite direction of circular polarization,
the polaritonic branches are clearly observed for negative $B$ and are strongly suppressed for positive $B$.
Finally, for linear polarized MWs polaritonic branches of nearly equal intensity are observed for both directions of
the field $B$.

\begin{figure}[tbp]
\begin{center}
\includegraphics[width=8.5cm]{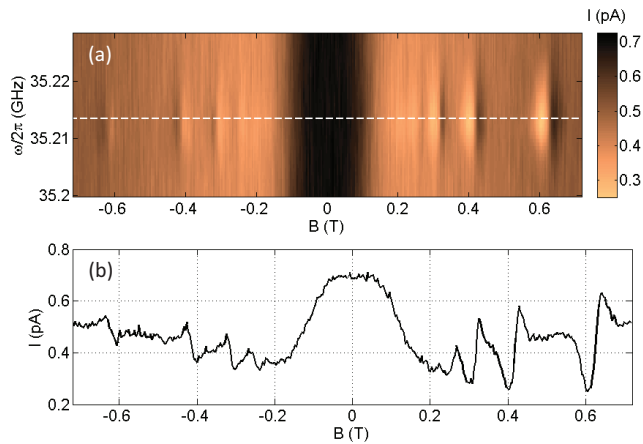}
\end{center}
\caption{(color online) (a) Photocurrent response of the 2D electron system at $T=0.2$~K and electron density  $n_s=5.2\times 10^6$~cm$^{-2}$ to circular polarized MWs plotted as a function of MW frequency $\omega$ and the perpendicular magnetic field $B$. (b) Photocurrent response taken at $\omega=\omega_r$ (indicated by a dashed line in panel (a)).}
\label{fig:2}
\end{figure}

Next, we consider the photoconductivity response of electrons at the harmonics of the CR by varying the value
of $B$ in a wide range for both directions of the field. Figure~\ref{fig:2}(a) shows the current signal
measured at $T=0.2$~K for MW excitation which is circularly polarized in the same direction as for the
data shown in Fig.~\ref{fig:1}. Strong photocurrent response is observed near values of $B$ which
satisfy the condition $\omega=m\omega_c$, where $m=2,3,...$ . In addition, the photocurrent response
is observed only when the frequency of microwave excitation $\omega$ is in the vicinity of the resonant
frequency $\omega_r$. In particular, Figure~\ref{fig:2}(b) displays the current signal measured at
$\omega=\omega_r$. As shown previously~\cite{YamAbdBad-2015}, the photoconductivity oscillations can
be observed only when the amplitude of the microwave $E$-field is sufficiently large, which occurs near the
cavity resonance. For the input MW power used in Fig.~\ref{fig:2} we crudely estimate a maximum amplitude
of the $E$-field in the resonant TEM$_{002}$ mode of about 5~V/cm.

The most important feature of plots shown in Figs.~\ref{fig:2}(a,b) is a significant asymmetry in the amplitude of the photocurrent response with respect to directions of $B$. In particular, the large current oscillations are observed in the direction of the $B$-field corresponding to strong CR absorption, while the oscillations are strongly reduced for the direction of the field corresponding to suppressed CR absorption, c.f. Fig.~\ref{fig:1}(a) and \ref{fig:1}(b). For the opposite direction of circular polarization (data not shown), the situation is reversed with respect to the direction of $B$, which is consistent with the behavior of CR absorption described above. Finally, for linear polarized microwaves, the amplitude of oscillations was found to be the same for both directions of $B$.

\begin{figure}[tbp]
\begin{center}
\includegraphics[width=9.75cm]{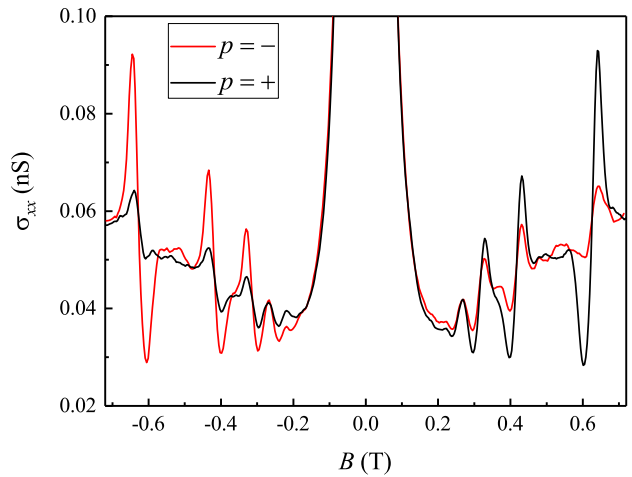}
\end{center}
\caption{(color online) Longitudinal conductivity $\sigma_{xx}$ of 2D electrons at $T=0.2$~K and $n_s=5.1\times 10^6$~cm$^{-2}$ versus magnetic field $B$ for two directions of circular polarization of MWs at frequency $\omega/2\pi=35.213$~GHz. A black line is for the circular polarization direction ($p=+ $) the same as for data shown in Figs.~\ref{fig:1} and \ref{fig:2}, while a red line is for the opposite direction ($p=-$).}
\label{fig:3}
\end{figure}

Figure~\ref{fig:3} shows values of $\sigma_{xx}$ extracted from the measured current signals and plotted as a
function of the magnetic field $B$ for two opposite directions of circular polarization. The strong dependence of the
amplitude of oscillations on the direction of circular polarization indicated in Fig.~\ref{fig:3} is the central result of this work and is at least in qualitative agreement with predictions of the theories based on photon-assisted scattering. For the sake of a quantitative comparison, we consider theoretical predictions with some more details below.

To obtain probabilities of photon-assisted scattering there is a
nonperturbative treatment resulting in the Landau-Floquet states which
includes the MW field in an exact form (for a recent examples, see~\cite%
{Par-2004,TorKun-2005,Mon-2017}). The wave function of these states has a time-dependent
shift of the center position $\xi \left( t\right) $ and a special phase
factor. Considering two components of the MW field $E^{\left( x\right)
}=a_{p}E_{\mathrm{mw}}\cos \omega t$, and $E^{\left( y\right) }=b_{p}E_{%
\mathrm{mw}}\sin \omega t$ (here the numbers $a_{p}$ and $b_{p}$ describe
the MW polarization, and $p$ is the polarization index) one have to introduce two time-dependent shifts $\xi $
(along $x-$axis) and $\zeta $ (along $y-$axis), which satisfy
classical equations of motion~\cite{Mon-2017}. The important consequence of
this treatment is that the new expression for the matrix elements $%
\left\langle n^{\prime },X^{\prime }\right\vert e^{-i\mathbf{q}\cdot \mathbf{%
r}}\left\vert n,X\right\rangle _{L-F}$ describing electron scattering off
disorder acquires an additional exponential factor%
\begin{equation}
\exp \left[ -i\beta _{p,\mathbf{q}}\sin \left( \omega t+\gamma _{p}\right) %
\right] ,  \label{e1}
\end{equation}%
as compared to the conventional form obtained for the Landau basis. Here%
\begin{equation}
\beta _{p,\mathbf{q}}=\lambda \omega _{c}l_{B}\frac{\sqrt{q_{y}^{2}\left(
a_{p}\omega _{c}+b_{p}\omega \right) ^{2}+q_{x}^{2}\left( a_{p}\omega
+b_{p}\omega _{c}\right) ^{2}}}{\left( \omega ^{2}-\omega _{c}^{2}\right) },
\label{e2}
\end{equation}%
$\lambda =eE_{\mathrm{mw}}l_{B}/\hbar \omega $ is the parameter describing the
strength of the MW field, $l_{B}^{2}=\hbar c/eB$, and the exact form of $%
\gamma _{p}$ is not important for the following discussion.

Calculation of scattering probabilities is reduced to the usual
treatment by means of the Jacobi-Anger expansion $e^{iz\sin \phi
}=\sum_{k}J_{k}\left( z\right) e^{ik\phi }$ [here $J_{k}\left( z\right) $ is
the Bessel function]. Thus, the probability of one-photon assisted
scattering is proportional to $J_{1}^{2}\left( \beta _{p,\mathbf{q}%
}\right) \simeq \beta _{p,\mathbf{q}}^{2}/4$, if the parameter $\beta _{p,%
\mathbf{q}}$ is small. Therefore, the ratio of MW-induced corrections to the dc
dissipative conductivity obtained for different directions of circular
polarization ($p=+$ and $p=-$; $a_{\pm }=1$, $b_{\pm }=\pm 1$) is described
by the simple relationship
\begin{equation}
\frac{\Delta \sigma _{xx}^{\left( +\right) }}{\Delta \sigma _{xx}^{\left(
-\right) }}=\frac{\left( \omega /\omega _{c} +1\right) ^{2}}{\left( \omega /\omega _{c}
-1\right) ^{2}}.  \label{e3}
\end{equation}%
This equation is valid for the both DM and IM because $\beta _{p,\mathbf{q}}$ is now
independent of the direction of the wave-vector $\mathbf{q}$. The ratio of Eq.~(\ref{e3}) is large for $m=2$ (it equals 9) and $m=3$
(it equals 4), but it approaches unity if $m$ increases. Calculations based on Eq.~(\ref{e2}) indicate that deviations from circularity
(elliptic polarization) can reduce the ratio $\Delta \sigma _{xx}^{\left( +\right) }/\Delta \sigma _{xx}^{\left(
-\right) }$.

\begin{figure}[tbp]
\begin{center}
\includegraphics[width=9.5cm]{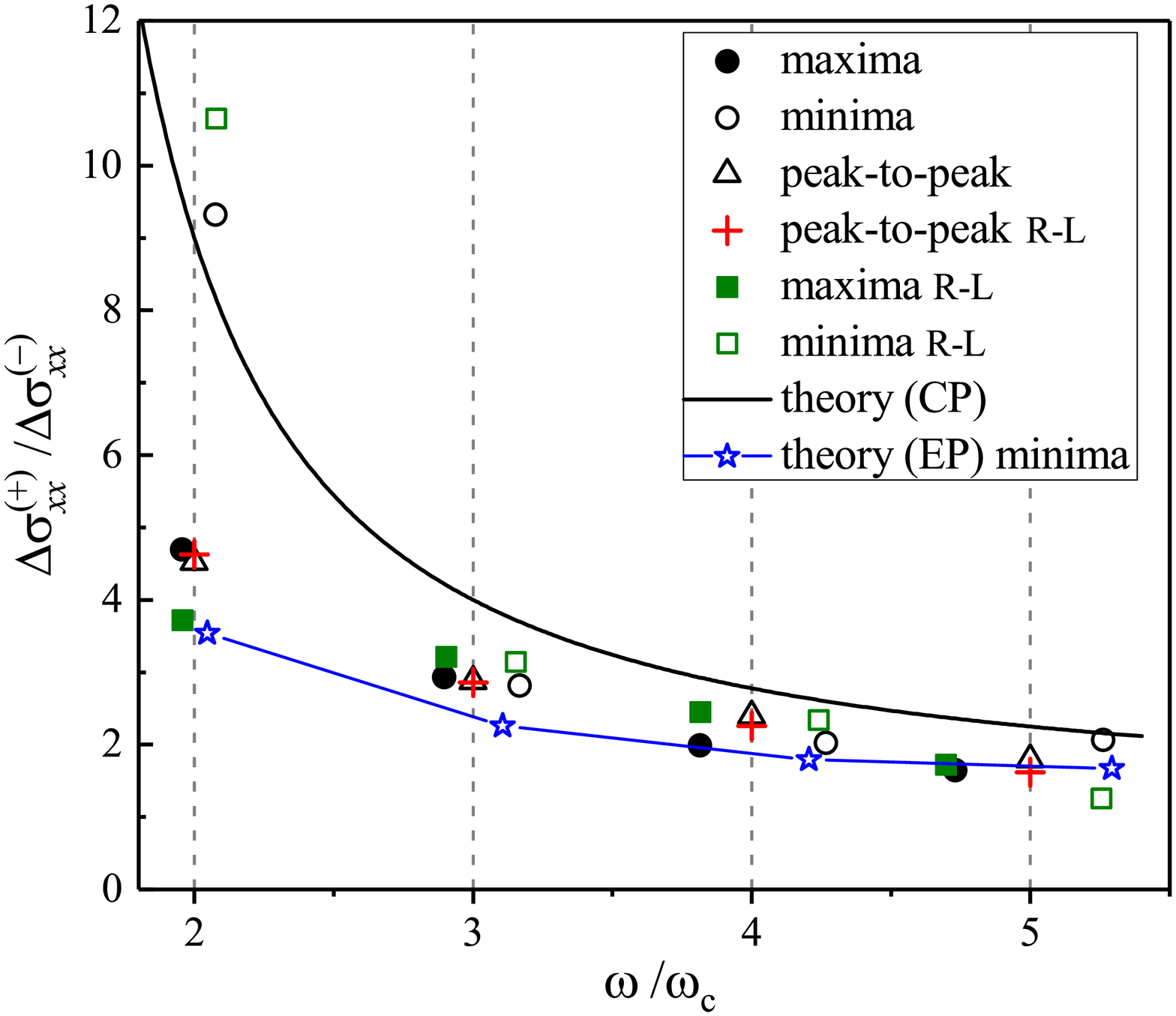}
\end{center}
\caption{(color online) The ratio $\Delta \sigma _{xx}^{\left( +\right) }/\Delta \sigma _{xx}^{\left(
-\right) }$ versus $\omega / \omega _{c}$: results obtained from experimental data for conductivity maxima (filled circles and squares) and
minima (open circles and squares), the ratio of peak-to-peak amplitudes (triangles and crosses), the ratio of amplitudes obtained for positive and negative $B$ using the same curve of Fig.~\ref{fig:3} are marked in the legend by R-L,
theory [Eq.~(\ref{e3})] calculated for the circular polarization (CP, solid curve), theory using the elliptic polarization (EP) with
$a_{\pm }=0.7$ and $b_{\pm }=\pm 1.3$ calculated for minima (open stars).}
\label{fig:4}
\end{figure}

For the both DM and IM, the shape of MIRO was shown to be similar to experimental observations~\cite{YamAbdBad-2015,Mon-2017}.
Therefore, here we concentrate on the comparison of polarization dependencies of MIRO. The
ratio of the amplitudes of conductivity oscillations obtained here for two opposite directions of circular
polarization is illustrated in Fig.~\ref{fig:4}. The experimental results were plotted
separately for maxima (filled circles and squares) and minima (open circles and squares) because the ratio
of Eq.~(\ref{e3}) shown in this figure by the solid curve depends strongly on $B$.
The ratio of peak-to-peak amplitudes shown by triangles agrees with these data.
To ensure that a possible power difference didn't affect our results here we plotted (squares and crosses) the ratio of respective
amplitudes obtained at positive and negative $B$ from the same curve of Fig.~\ref{fig:3}.

Fig.~\ref{fig:4} indicates that the ratio $\Delta \sigma _{xx}^{\left( +\right) }/\Delta \sigma _{xx}^{\left(
-\right) }$ found in the experiment
increases with lowering $\omega / \omega _{c}$ in accordance with the theory, still
it at an average is lower than theoretical values by a factor of the order of unity (about $1.4$).
As a possible reason for such a reduction we considered a deviation from circularity.
By way of illustration, calculations employing the DM and Eq.~(\ref{e2}) with $a_{\pm }=0.7$ and $b_{\pm }=\pm 1.3$ are shown in
Fig.~\ref{fig:4} by star symbols with lines. Actually, the chosen parameters of ellipticity give even stronger reduction of
the ratio $\Delta \sigma _{xx}^{\left( +\right) }/\Delta \sigma _{xx}^{\left(
-\right) }$ than that observed in the experiment. It should be noted that chosen ellipticity still leads to a strong suppression (about 10 times) of the photocurrent response at CR conditions for MWs with $p=-$, which is consistent with observations shown in Fig.~\ref{fig:1}.  For inverted parameters ($a_{\pm }=1.3$, $b_{\pm }=\pm 0.7$),
the reduction of the ratio of amplitudes is less than 5\% because the integrand of the conductivity equation contains the symmetry-breaking factor
$q_{y}^{2}$. Since the Corbino experiment gives $\sigma _{xx}$ data averaged over all directions of the driving electric field, the actual reduction
of the ratio $\Delta \sigma _{xx}^{\left( +\right) }/\Delta \sigma _{xx}^{\left(
-\right) }$ for $a_{\pm }=0.7$ and $b_{\pm }=\pm 1.3$ can be noticeably less than that shown in Fig.~\ref{fig:4} by star symbols. 

It should be noted also that under the conditions of the experiment the
average Coulomb interaction energy per electron is more than 30 times larger than the average kinetic
energy (temperature). Additionally, one cannot completely exclude heating of surface electrons by MWs.
Therefore, the agreement between the experiment and the theories based on photon-assisted scattering
can be considered as satisfactory.

At $E_{\mathrm{mw}}=5\,\mathrm{V/cm}$, the both DM and IM are estimated to provide sufficient amplitudes of MIRO under the conditions of this
experiment, but the circular-polarization study cannot determine the contribution of which model dominates.
For liquid $^{3}\mathrm{He}$, there is also an uncertainty in the definition of the inelastic thermalization
rate because short wavelength ripplons responsible for inter-level scattering are heavily damped at low temperatures. Assuming that we can still rely on the usual ripplon spectrum, the IM yields somewhat stronger amplitudes of MIRO than the DM. Still, for a strict conclusion, an experimental setup with different linear polarizations and a fixed direction of the dc field is required.

In summary, by irradiating a nondegenerate 2D electron gas formed on liquid $^{3}\mathrm{He}$ with MWs of
different polarizations we discovered a strong dependence of MIRO on the MW circular polarization direction.
This allowed us to report the first observation of the effect of radiation helicity which provides a crucial
information for understanding the origin of MW-induced magnetoconductivity oscillations in a 2D electron gas.
In particular, our experiments unambiguously support theoretical mechanisms of MIRO based on photon-assisted
scattering off disorder.

This work is supported by an internal grant from Okinawa Institute of Science and Technology (OIST) Graduate University. We are grateful to V.~P. Dvornichenko for providing technical support.

\end{document}